# Close-Coupling Time-Dependent Quantum Dynamics Study of the H + HCl Reaction

### Li Yao,[†,‡] Ke-Li Han,*,[†] He-Shan Song,*,[‡] and Dong-Hui Zhang[§]

*Center for Computational Chemistry and State Key Laboratory of Molecular Reaction Dynamics, Dalian Institute of Chemical Physics, Chinese Academy of Sciences, Dalian 116023, China, Department of Physics, Dalian University of Technology, Dalian 116023, China, and Department of Computational Science, National University of Singapore, Singapore*

The paper presents a theoretical study of the dynamics of the H + HCl system on the potential energy surface (PES) of Bian and Werner (Bian, W.; Werner, H. -J., J. Chem. Phys. 2000, 112, 220). A time-dependent wave packet approach was employed to calculate state-to-state reaction probabilities for the exchanged and abstraction channels. The most recent PES for the system has been used in the calculations. Reaction probabilities have also been calculated for several values of the total angular momentum $J > 0$. Those have then been used to estimate cross sections and rate constants for both channels. The calculated cross sections can be compared with the results of previous quasiclassical trajectory calculations and reaction dynamics experimental on the abstraction channel. In addition, the calculated rate constants are in the reasonably good agreement with experimental measurement.

## 1. Introduction

The gas-phase reaction of H + HCl and the corresponding reverse reaction of Cl atoms with molecular H represent important elementary steps in the $H_2 + Cl_2 \rightarrow 2HCl$ reaction system, which has played a major role in the development of chemical kinetics and to the environment in atmospheric chemistry.[1,2] A large number of kinetics studies were carried out for the H + HCl elementary reaction in the temperature range 195 K $\leq T \leq$ 1200 K,[3] including experiments in which the influence of selective vibrational excitation of HCl on the reaction rate was investigated.[4−8] Recent studies[9−16] of the reaction H + HCl at a collision energy of 1.6 eV have measured the integral cross section for the abstraction channel to be $(2\pm1)$ Å$^2$,[9,10] where as the exchange plus energy-transfer channels gave a combined cross section of $(13\pm3)$ Å$^2$.[9,10] It has been shown by Wight et al.[11] that the dominant process is energy transfer.[2] In contrast to numerous experiments, only a few theoretical studies have been reported for this H + HCl reaction.

Due to the relatively low collision energies employed in the Cl + H$_2$ dynamics experiments, these studies provided detailed information about the region of the HClH potential energy surface (PES) close to the threshold of reaction.[17−23] However, much less information is available concerning the high-energy regions of the PES that can be assessed in H + HCl dynamics experiments using translationally excited H atoms.[24−27] The first globally realistic Cl−H−H PES was calculated by Baer and Last,[26] and a more recent PES was published by Truhlar et al.[27] Therefore, a scaled PES has been computed by Bian and Werner (BW2)[28] to get the dissociation energies right. The H$_2$Cl reaction system was presented by Bian et al.[28] more recently. The PES was developed using the highly accurate electronic structure methods based on extensive ab initio calculations and very large basis sets presently applied. The ab initio calculations were carried out at more than 1200 nuclear geometries.[28] The later version of the BW2 PES was further improved by scaling the correlation energies at all geometries with a constant factor.[28]

A major theoretical problem stems from the inability of current ab initio and semiempirical methods in providing a reliable PES. Recently, Aoiz et al.[29,30] have carried out quasiclassical trajectory (QCT) calculations on two versions of the BW2 PES for the reaction.[30] The calculated cross sections are in reasonably good agreement with experimental measurement. This provides some assurance that the BW2 PES may be reasonably accurate near the transition state. To the best of our knowledge, no time-dependent wave packet (TDWP) study for the reaction system corresponding to a collision energy range of [0.1,1.4]eV was reported on the BW2 PES.

In the present work, the reaction probabilities have been calculated by employing TDWP method for several values of the total angular momentum quantum number $J > 0$ on BW2 PES. Those have then been used to estimate cross sections and rate constants for both channels. The calculated cross sections can compared with the results of previous QCT calculations and reaction dynamics experiments of Aoiz et al.[30] and experiments of Brownsword et al.[25] on the abstraction channel. The thermal rate constants of HCl are calculated by employing the uniform $J$-shifting method,[32,33] and a comparison with experimental measurement is provided.

In this paper, we have carried out TDWP calculations for the following two channels of the H + HCl reaction to figure out the different dynamical properties of the I$_A$ abstraction channel and the I$_E$ exchange channel.

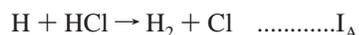

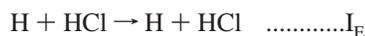

This paper is organized as follows: Section 2 gives a brief review of the theoretical methodologies used in the current study. The result of numerical calculation and discussion of the result are given in Section 3. Comparisons with other theoretical calculations and with experimental measurement, whenever possible, are also given in this section. Section 4 concludes the article.

---

* Corresponding authors: E-mail: klhan@ms.dicp.ac.cn.
† Dalian Institute of Chemical Physics, Chinese Academy of Sciences.
‡ Dalian University of Technology.
§ National University of Singapore.

## 2. Theory

Here, we briefly describe the TDWP method employed to calculate the initial state-selected total reaction probability with the final resolved products. The reader is referred to ref 34 for more detailed discussions of the methodology. In the present study, we solve the time-dependent Schrödinger equation

$$i\hbar \frac{\partial \Psi}{\partial t} = H\Psi \quad (1)$$

for the H + HCl reaction. The Hamiltonian expressed in the reactant Jacobi coordinates for a given total angular momentum quantum number $J$ can be written as

$$H = -\frac{\hbar^2}{2\mu_R}\frac{\partial^2}{\partial R^2} + \frac{(\vec{J}-\vec{j})^2}{2\mu_R R^2} + \frac{\vec{j}^2}{2\mu_r r^2} + V(\vec{r},\vec{R}) + h(r) \quad (2)$$

where $\vec{r}$ is the diatom internuclear, $\vec{R}$ is the vector jointing the center of mass of diatom to the atom, while $\mu_r$ is the reduced mass for HCl, and $\mu_R$ is the reduced mass between H and HCl. $\vec{J}$ and $\vec{j}$ represent the total angular momentum operator and the rotational angular momentum operator of HCl, respectively. $V(\vec{r},\vec{R})$ is the interaction potential excluding the diatomic potential of the diatom. The diatomic reference Hamiltonian $h(r)$ is defined as

$$h(r) = -\frac{\hbar^2}{2\mu_r}\frac{\partial^2}{\partial r^2} + V_r(r) \quad (3)$$

where $V_r(r)$ is a diatomic reference potential.

The time-dependent wave function satisfying the Schrödinger eq 1 can be expanded in terms of the body-fixed translational−vibrational−rotational basis, defined using the reactant Jacobi coordinates, as[35]

$$\Psi_{v_0 j_0 K_0}^{JM\epsilon}(\vec{R},\vec{r},t) = \sum_{n,v,j,K} F_{nvjK,v_0 j_0 K_0}^{JM\epsilon}(t) u_n^v(R)\phi_v(r) Y_{jK}^{JM\epsilon}(\hat{R},\hat{r}) \quad (4)$$

where $n$ is the translational basis label and $M$ and $K$ are the projection quantum numbers of $J$ on the space-fixed $z$ axis and body-fixed $z$ axis, respectively. $(v_0, j_0, K_0)$ denotes the initial rovibrational state, and $\epsilon$ is the parity of the system defined as $\epsilon = (-1)^{j+L}$ with $L$ being the orbital angular momentum quantum number. The reader can find the definitions of various basis functions elsewhere.[34]

The split-operator method[36] is employed to carry out the wave packet propagation. The time-dependent wave function is absorbed at the edges of the grid area to avoid artificial reflections.[37] Finally the initial state-selected total (final-state-summed) reaction probabilities are obtained through the flux calculation[35] at the end of the propagation.

$$P_{v_0 j_0 K_0}^{J}(E) = \\ \langle \psi_{v_0 j_0 K_0}^{JM\epsilon+}(E)|^1/_2[\delta(\hat{s}-s_0)\hat{v}_s + \hat{v}_s \delta(\hat{s}-s_0)]|\psi_{v_0 j_0 K_0}^{JM\epsilon+}(E)\rangle \quad (5)$$

where $s$ is the coordinate perpendicular to a surface located at $s_0$ for flux evaluation and $v_s$ is the velocity operator corresponding to the coordinate $s$. $\psi_{v_0 j_0 K_0}^{JM\epsilon+}(E)$ is the time-independent wave function that can be obtained by Fourier transforming the TDWP wave function.

Once the reaction probabilities $P_{v_0 j_0 K_0}^{J}(E)$ have been calculated for all fixed angular momenta $J$, calculation for the cross sections and rate constants are straightforward. The cross section is given by

$$\sigma_{v_0 j_0}(E) = \frac{\pi}{k_{v_0 j_0}^2}\sum_J (2J+1) P_{v_0 j_0}^{J}(E) \quad (6)$$

where $k_{v_0 j_0} = (2\mu_R E)^{1/2}/\hbar$ is the wavenumber corresponding to the initial state at fixed collision energy $E$, and $P_{v_0 j_0}^{J}(E)$ is given by

$$P_{v_0 j_0}^{J}(E) = \frac{1}{2j_0+1}\sum_{K_0} P_{v_0 j_0 K_0}^{J}(E) \quad (7)$$

In practice, we can use the interpolation method to get the probabilities for missing values of $J$; reaction probabilities at only a limited number of total angular momentum values of $J$ need to be explicitly calculated.

As in refs 34, 38, we construct wave packets and propagate them to calculate the reaction probabilities for $P^J(E)$ each product. The integral cross section from a specific initial state $j_0$ is obtained by summing the reaction probabilities over all the partial waves on total angular momentum.

The rate constants are calculated for the initial states ($v = 0$, $j = 0$) of HCl by using the uniform version[32] of $J$-shifting approximation.[33] The initial state-specific thermal rate constant in the uniform $J$-shifting scheme is given as

$$K(T) = \sqrt{\frac{2\pi}{(\mu_R k_B T)^3}} Q^0(T) \sum_J (2J+1) e^{-B_J(T)J(J+1)/k_B T} \quad (8)$$

The shifting constant is determined by[32]

$$B_J(T) = \frac{k_B T}{J(J+1) - J_i(J_i+1)} \ln\left(\frac{Q^{J_i}}{Q^J}\right) \quad (9)$$

where $k_B$ is the Boltzmann constant, $T$ is the temperature, and $Q^{J_i}$ is a partitionlike function defined as

$$Q^{J_i} = \int P^{J_i}(E) e^{-E/k_B T} dE \quad (10)$$

where $J_i$ is a reference angular momentum that divides total angular momentum into different ranges,[32] and $Q^J$ is similarly defined as

$$Q^J = \int P^J(E) e^{-E/k_B T} dE \quad (11)$$

where $P^J(E)$ is the probabilities for a total angular momentum from a given initial state.

The numerical parameters for the wave packet propagation are as follows: A total number of 200 sine functions (among them 80 for the interaction region) are employed for the translational coordinate $R$ in a range of $[0.8, 14.0]a_0$. A total of 100 vibrational functions are employed for $r$ in the range of $[0.8, 8.5]a_0$ for the reagents HCl in the interaction region. For the rotational basis, we use $j_{max} = 45$. The number of $K$ used in our calculation is given by $K_{max} = \max(3, K_0+2)$ starting with $K_0 = 0$. The largest number of $K$ used is equal to 6 for the $j = 0$, $K_0 = 0$ initial state (for $\epsilon = -1$, there is one less $K$ block used). These values of $K_0$ and $K_{max}$ were determined following an extensive series of tests.[35] It was found that convergence of total cross sections, for all the reported initial (rotational) states in the entire energy region, was achieved up to a few percent. The initial wave packet was centered at $R = 10a_0$, with a width of $0.23a_0$ and an average translational energy of 0.8 eV.[35] For lower $J$, we propagate the wave packets for 15000 au of time to converge the low energy reaction probability (in all calculations, a time step-size of 10 au was used). For $J > 20$, we propagate the wave packets for a shorter time, because the reaction probability in the low energy region is negligible.[35] In this calculation, we used $J$ from 0 to 80 to calculate the cross section.

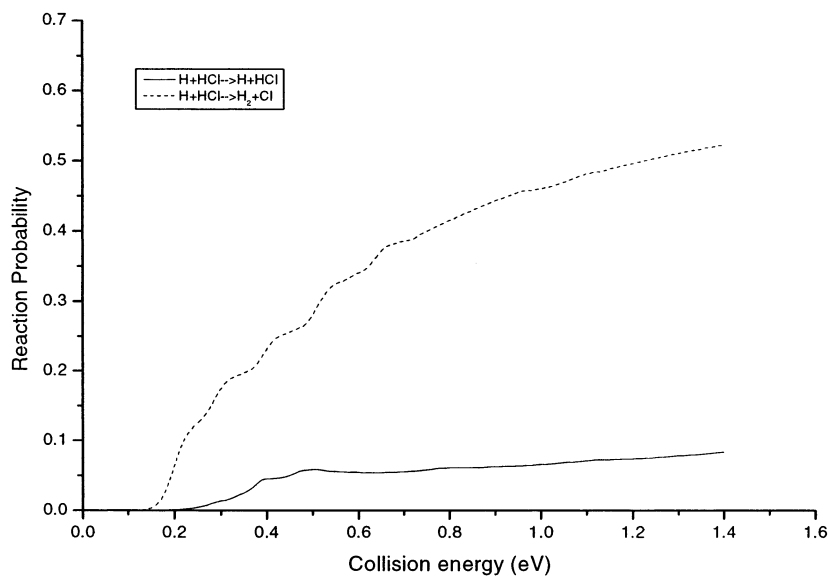

**Figure 1.** Total reaction probabilities for $J = 0$ from the ground state of the HCl reactant for both channels of the H + HCl reaction on the BW2 PES. The solid line is for the exchange channel, and the dashed line is for the abstraction channel.

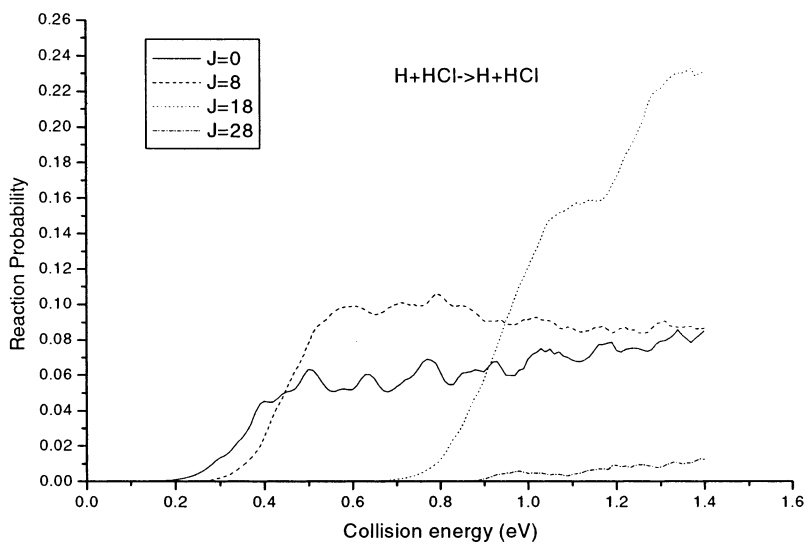

**Figure 2.** Total reaction probabilities for $v = 0$, $j = 0$, $J = 0, 8, 18, 28$ of the HCl reactant for the exchange channel on the BW2 potential. The solid line is for $J = 0$, the dashed line is for $J = 8$, the dotted line is for $J = 18$, and the dashed−dotted line is for $J = 28$.

### 3. Results and Discussion

**Reaction Probabilities.** First of all, we computed the energy resolved reaction probabilities for collision energies in the range of [0.1,1.4]eV with HCl initially in its ground state. The results of $J = 0$ as a function of the collision energy for the BW2 potential for all possible channels are presented in Figure 1. As shown in Figure 1, the behavior of the reaction probabilities for the two channels is quite different. One can see from Figure 1 that the reaction probability for the abstraction channel is much higher than that of the exchange channel.

In the entrance channel of the H + HCl → H$_2$+Cl reaction, a van der Waals (vdW) well with a collinear geometry and a depth of 0.019 eV is found, while in the exit channel, a T-shaped vdW well with a depth of 0.022 eV is predicted.[28] In the collinear transition state for the abstraction reaction, the heights of the classical barriers is 0.184 eV for BW2 PES.[28] For the H + ClH exchange reaction, which also has a collinear transition state, the barrier height is computed to be 0.776 eV for BW2 PES.[28] The threshold of the reaction for the exchange channel is a bit higher than that of the abstraction channel. This can explain why the two channels show different behavior in Figure 1.

In addition, we calculated the reaction probabilities for different total angular moment $J$ for HCl initially in its ground state. The reaction probabilities as a function of collision energy for total angular momentum of $J = 0, 8, 18$ and 28, are presented for both channels in Figure 2 and 3. As shown in Figure 2, generally the values of the reaction probabilities decrease with an increase of $J$ in the low energy region. However, this is not always true. For example, the $J = 8, 18$ reaction probabilities exceed the $J = 0$ probability at high energies (>0.45 and

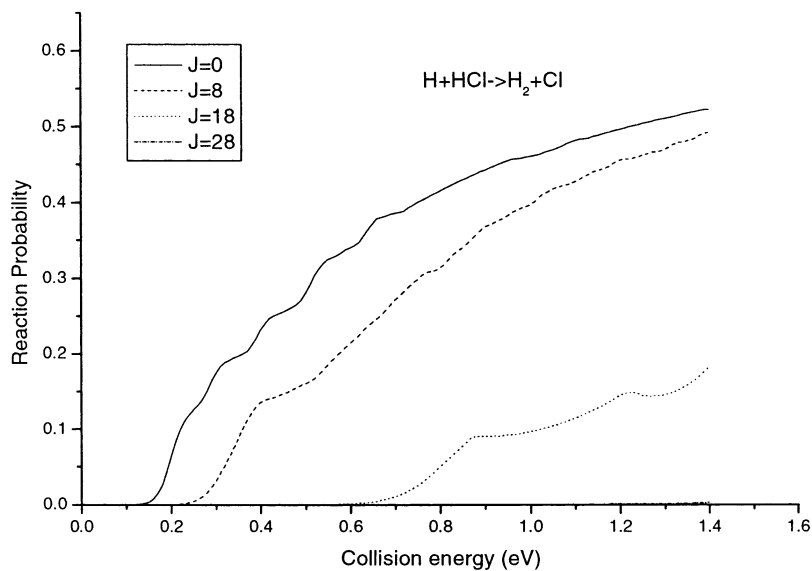

**Figure 3.** Total reaction probabilities for $v = 0$, $j = 0$, $J = 0, 8, 18, 28$ of the HCl reactant for the abstraction channel on the BW2 potential. The solid line is for $J = 0$, the dashed line is for $J = 8$, the dotted line is for $J = 18$, and the dashed−dotted line is for $J = 28$.

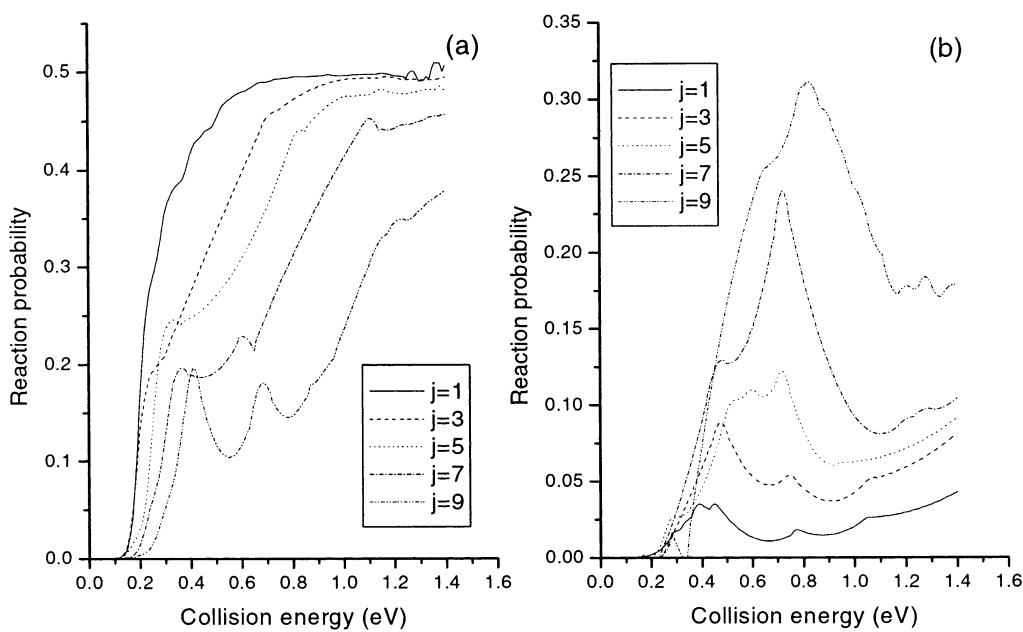

**Figure 4.** Total reaction probabilities for $J = 0$, $v = 0$ of the HCl reactant for H + HCl → H + HCl (a) H + HCl → $H_2$ + Cl (b) on the BW2 potential. The solid line is for $j = 1$, dashed line is for $j = 3$, dotted line is for $j = 5$, dashed−dotted line is for $j = 7$, dashed−dotted−dotted line is for $j = 9$.

>0.9 eV), see Figure 2, while the values of the reaction probabilities for the abstraction channel decrease with an increase of $J$ as shown in Figure 3. The threshold of the probabilities increases with increasing $J$ in both channels.

The effect of the initial reagent rotation excitation on the reaction probability for the two channels ($J = 0$, $v = 0$, $j = 1$, 3, 5, 7, 9) is shown in Figure 4. As seen from Figure 4, the reaction probabilities of the abstraction channel decrease with increasing rotational quantum number $j$. The decrease may be due to shape resonances caused by hydrogen tunneling through a centrifugal barrier, which traps the hydrogen for a finite time. The reaction probabilities for the exchange channel increase with increasing $J$. The increase and the oscillating can perhaps be explained by a long-range vdW well in both the entrance and exit channels that is the same as the explanation in ref 12. The oscillating of the probabilities turns stronger as $j$ increases for the exchange channel.[31] The negative values of the reaction probabilities in the low energy region is negligible with the approximate of the theory.

**Integral Cross Sections.** Next, we calculate the integral cross section from the initial ground state of HCl on the BW2 surface. In ref 30, Aoiz and Bañares carried out QCT calculation of the reaction cross section on the BW2 PES with HCl initially in its state ($v = 0$, $j = 0-6$). The calculated cross sections for H + HCl are depicted in Figure 5 for both channels. As can be seen, the present cross section in the abstraction channel is systematically larger than the results of QCT[30] and the corresponding experimental values. For the exchange channel, the results from the QCT calculations cross with the ones from the TDWP calculations at 1.2 eV, and the threshold energy of the QCT results is much higher than the one from this work. The threshold energy of the TDWP results is much lower than the reaction barrier for this channel (0.8 eV). This can be explained by the fact that the TDWP method has correctly included the zero-point energy and tunneling effects. Total absolute reaction cross sections for chlorine atom information, $\sigma_R(1.4\ eV) = (0.35\pm0.16)$ Å$^2$, have been measured using a photolytic calibration method.[30] In ref 25, a value of $\sigma_R(1.0) = (0.34\pm0.05)$ Å$^2$ was determined employing vacuum-UV laser-induced fluorescence for Cl atom detection. However, in any case, the agreement is only approximate, bacause the BW2 cross sections lie outside the experimental points. The experimental values of $(2\pm1)$ Å$^2$ reported in ref 9 for Ecol = 1.6 eV was not included in Figure 5.

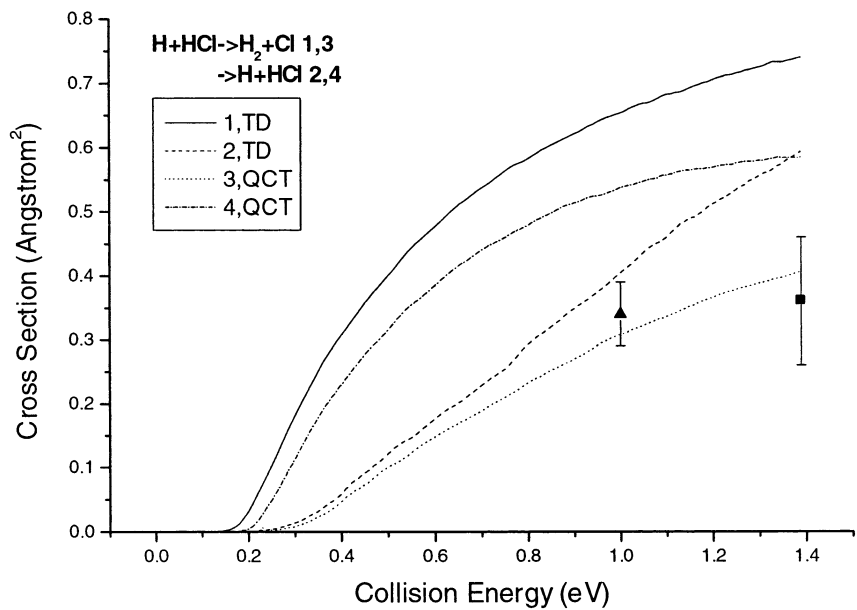

**Figure 5.** Reaction Cross section as a function of collision energy for the H + HCl reaction using TDWP approach and QCT on the BW2 PES. The results of the experimental measurements filled triangle of ref 36 and filled square of ref 16 as well as QCT excitation function calculated on the BW2 PES are also shown for comparison of the abstraction channel. The solid line and the dashed line are separately calculated for the abstraction channel and the exchange channel using the TDWP approach. The dotted line and the dashed−dotted line are in the results for the abstraction channel and exchange channel from QCT calculations. No experimental cross sections have been reported for the exchanged channel of the reaction.

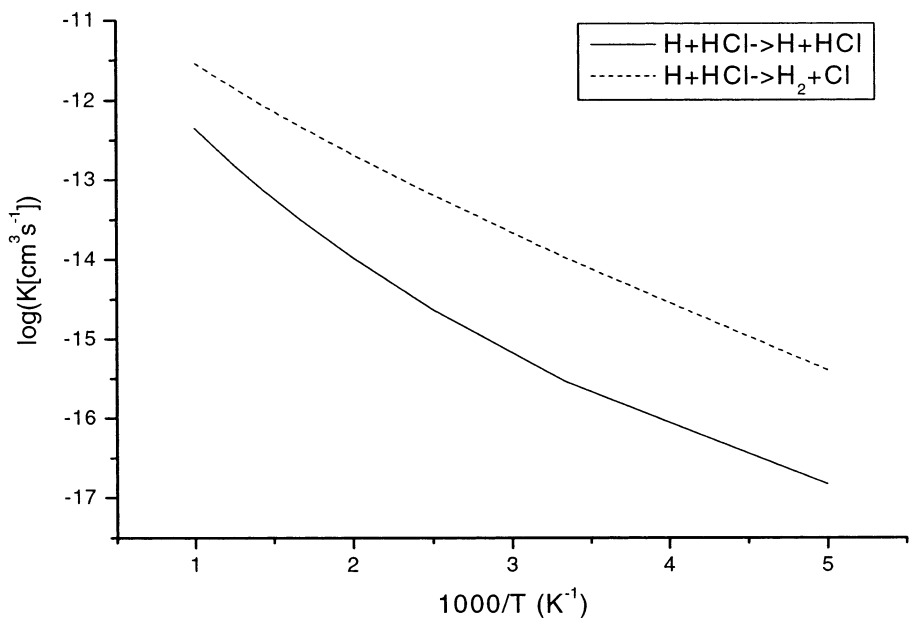

**Figure 6.** The Rate Constant for $v = 0$, $j = 0$ of the HCl reactant for the H + HCl reaction eq I on the BW2 potential. The solid line represents the exchange channel, and the dashed line represents the abstraction channel.

At low temperature and collision energies, the tunneling works most remarkably in the reaction process to make the H abstraction easier. At high collision energies, the H atom has more chances to collide with the Cl atom, and the reaction will produce more HCl; therefore, the result for the exchange channel increases faster than the one for the abstraction channel. It is surprising that dynamical calculations on the BW2 PES, which is based on high-quality ab initio points and clearly more accurate than any previous PES,[28] yield reaction cross sections for the abstraction channel noticeably larger than the experimental ones.[30] No experimental cross sections have been reported for the exchanged channel of the reaction.[30]

**Rate Constant.** Despite several measurements of relative reaction rates, accurate results for absolute rate constants have dodged investigators for many years. So, the rate constant calculations are also one of our main objectives.

One has to calculate the total reaction probability for more than two values of $J$ to obtain a more accurate rate constant.[32]

A very accurate rate constant can be obtained by using reaction probabilities evaluated at more than 6 values of $J$ (partial waves). The HCl initially in its state ($v = 0, j = 0$) had been considered in the calculation of $K(T)$.

The calculated rate constants for the H + HCl reaction are depicted in Figure 6 for both channels. The calculation of the thermal rate constants is in the range of temperatures between 200 and 1000 K. One can see that the calculated rate constant of the abstraction channel is systematically larger than that of the exchange channel for all the temperatures. As can be seen, the present rate constants roughly follow the same as the corresponding experimental values in the range of $[10^{-17}, 10^{-12}]$ cm$^3$s$^{-1}$.[2,26,27,39–41] However, in any case, the agreement is only approximate, because the rate constants on BW2 PES lie inside the range of experimental values. The agreement of the results is presented here, and the experimental results are good enough, but the BW2 PES need improvement of the theoretical results.

The calculated rate constant can also be taken as an indication that the existence of quantum effects such as tunneling effect may play an important role. So, according to our calculation results here, possible reasons for the nonlinear behavior of the Arrhenius plots of the reactions should be the combination influence of tunneling effects.

One can predict that the rate constants of the exchange channel from calculated reaction probabilities should also be smaller. This feature is usually found for exothermic reactions with a low energy barrier.[42,43]

### 4. Conclusions

In this work, we have applied the TDWP approach to study the H + HCl reaction on the BW2 PES. We have investigated the reaction probabilities as a function of the collision energy for the both channels for the H + HCl reaction and studied the influence of the initial rotational state excitation of the reagents. In the low temperature and collision energy range, the tunneling works most remarkably in the reaction process to make the H abstraction easier. In the high temperature and collision energy region, the H atom has more chances to collide with the Cl atom, and the reaction will produce more HCl. Thus, the cross section for exchange channel increases faster than that of abstraction channel. Such a study provides a clear and simple picture about reaction mechanisms.

**Acknowledgment.** This work is supported by NSFC (Grants Nos. 29825107 and 29853001) and NKBRSF.